\documentclass[floats,aps,showpacs,amssymb,prd,onecolumn,secnumarabic%
,tightenlines%
,nofootinbib]{revtex4}
\usepackage{amssymb}
\usepackage{graphics}
\usepackage{graphicx}
\usepackage{amsmath}
\usepackage{amsfonts}
\usepackage{bm}
\usepackage[colorlinks,linkcolor=red,anchorcolor=blue,citecolor=green]{hyperref}

\begin{document}

\title{Constraining quantum fluctuations of spacetime foam from BBN}

\author{S. Das$^1$}
\email[email: ]{saurya.das@uleth.ca}

\author{G. Lambiase$^{2,3}$}
\email[email: ]{lambiase@sa.infn.it}

\author{E.C. Vagenas$^{4}$}
\email[email: ]{elias.vagenas@ku.edu.kw}

\affiliation{
$^1$Theoretical Physics Group and Quantum Alberta, Department of Physics and Astronomy,
University of Lethbridge, 4401 University Drive, Lethbridge, Alberta, T1K 3M4, Canada.}

\affiliation{$^2$Dipartimento di Fisica ''E.R. Caianiello'' Universit\`a di Salerno, I-84084 Fisciano (Sa), Italy,}

\affiliation{$^3$INFN - Gruppo Collegato di Salerno, Italy.}

\affiliation{$^4$Theoretical Physics Group, Department of Physics, Kuwait University, P.O. Box 5969, Safat 13060, Kuwait}

\def\be{\begin{equation}}
\def\ee{\end{equation}}
\def\al{\alpha}
\def\bea{\begin{eqnarray}}
\def\eea{\end{eqnarray}}

\renewcommand{\theequation}{\thesection.\arabic{equation}}
%
%
%
\begin{abstract}
%
%
%
\par\noindent
A possibility to describe quantum gravitational fluctuations of the spacetime background is provided by virtual $D$-branes. These effects may induce a tiny violation of the Lorentz invariance (as well as a possible violation of the equivalence principle). In this framework, we study the formation of light elements in the early Universe (Big Bang Nucleosynthesis). 
By using the Big Bang Nucleosynthesis observations, We infer an upper bound on the topological fluctuations in the spacetime foam vacuum $\sigma^2$, given by $\sigma^2 \lesssim 10^{-22}$.
\end{abstract}

\pacs{04.50.-h, 04.60.Bc}

\maketitle

\section{Introduction}
\setcounter{equation}{0}
%
%
%
\par\noindent
Formulating a quantum theory of gravity is one of the most important challenges of the modern approaches aimed to unify all fundamental interactions. These studies have clearly
shown that spacetime must have a non-trivial topology near the Planck scale. After the Wheeler
\cite{wheeler} suggestion that
spacetime may have a foam-like
structure, the study of quantum
fluctuations of the spacetime background have received a lot of interest \cite{hawking80,hawking82,ellis84,ellis92,garay1, garay2,ellis99, ellis99_1}.

The Planck-size topological fluctuations imply that the
(quantum gravitational) vacuum behaves as a non-trivial medium.
This occurs, for example, in the framework of string theory
\cite{amelino97} and of the canonical approach to quantum gravity \cite{pullin,yu}. 
The underlying idea of Ref. \cite{amelino97} is that the quantum gravitational fluctuations in the vacuum get modified by the
passage of an energetic particle, inducing  recoil effects described by
back reaction effects on the propagating particle
\cite{ellis00}.
Although present technologies preclude any
possibility to probe quantum gravity effects, it has been suggested in Ref. \cite{amelino98} that Gamma-Ray
Bursts (GRBs) might offer the possibility to test the theories at Planck energies. The idea is that the origin of GRBs at a cosmological distance and their high energies may make them sensitive to the
dispersion scales that are comparable with the Planck scales \cite{amelino98}. In addition, the quantum fluctuations of spacetime may have had relevant consequences during the early Universe. In fact, the CPT-violating aspects of brane Universe models may induce an asymmetry between particles and antiparticles, allowing to explain the observed Baryon Asymmetry \cite{nickEPJC}.

In this contribution, we investigate the foamy
structure of the gravitational background, referring in particular to the
Ellis-Mavromatos-Nanopoulos-Volkov (EMNV) model \cite{ellis00a},
and its role on the formation of light elements during the primordial phase of the Universe evolution (Big Bang Nucleosynthesis).
Big-Bang Nucleosynthesis (BBN) represents an important epoch during the evolution of
the Universe. 
During this period, the primordial light 
elements formed leaving imprints on their abundance today.
Thanks to the advancements in measurements and theoretical predictions of the abundances of light elements, BBN 
has become a powerful cosmological probe for testing the early Universe. BBN has hence no trivial consequences on any physics scenario 
beyond the Standard Models of particle physics and cosmology \cite{Capozziello:2017bxm, Barrow:2020kug, Asimakis:2021yct}. The latter may alter the course of the events at that era with respect to the standard theories, and therefore such a probe provides strong constraints.

The rest of the paper is structured as follows. In section 2, we present the EMNV model and provide a formula that connects the baryon density parameter with the quantum fluctuations mass scale. In section 3, we provide bounds on the quantum fluctuations mass scale using today's primordial abundances of light elements which were produced in the BBN era. In section 4, we conclude and briefly present our results.
%
%
%
%
\section{The EMNV model}
\label{sec:2}
%
%
%
\par\noindent
The basic idea of the EMNV model is that the recoil effect of a $D$-brane struck by particles (bosons \cite{ellis99,ellis99_1,ellis00} or  fermions \cite{ellis00a}), induces an energy dependence of the off-diagonal terms of the background metric, $G_{0i}\sim u_i$, where $u_i\sim E/M_s\ll 1$ ($u_i$ is the average recoil velocity of the generic $D$-brane \cite{ellis00a} and
$E$ is the energy of the particle scattering off the $D$-brane, and $M_s$  characterizes the quantum fluctuations scale). The consequence of the off-diagonal term in the metric tensor implies the breaking of the Lorentz invariance \cite{ellis00a}.
For a $D$-dimensional spacetime, one has $G_{ij}=\delta_{ij}$, $G_{00}=-1$, and $G_{0i}\sim u_{i\,\parallel}$, $i, j=1, \ldots , D-1$ (here $u_{i\,\parallel}$ is the recoil (parallel) velocity of the $D$-particle). Moreover, the metric
induces a variation of the light velocity $\delta c/c\sim -E/M_s$.
The capture and splitting of the open string and its interaction with the $D$-particle, and the recoil of the latter, gives rise to a local
effective spacetime metric distortion \cite{ellis99, ellis99_1,Bernabeu:2006av}
\begin{equation}
ds^{2}=g_{\mu\nu}dx^{\mu}dx^{\nu}=(\eta_{\mu\nu}+G_{\mu\nu})dx^{\mu}dx^{\nu}~.\label{recmetric}
\end{equation}
The dispersion relation of a particle (neutrino)  propagating on the  deformed isotropic spacetime reads $g_{\mu\nu} p^{\mu}p^{\nu}= (\eta_{\mu\nu}+G_{\mu\nu}) p^{\mu}p^{\nu}=-m^{2}\,\Rightarrow\, E^{2}-2E{\vec{p}}\cdot u_{\parallel}-{\vec{p}}^{2}-m^{2}=0$, where $m$ is the mass of the particle. 
Taking into consideration this on-shell condition  and taking the average $\ll\dots\gg$ over $D$-particle populations with the stochastic processes ($\ll u_{i\,\parallel}\gg=0$, $\ll u_{i\,\parallel}u_{j\,\parallel}\gg=\sigma^{2}\delta_{ij}$), one gets the average neutrino and anti-neutrino energies in the $D$-foam background
\begin{equation}
\ll E_{\nu, \overline{\nu}}\gg  =  \sqrt{p^{2}+m_{\nu}^{2}}\left(1+\frac{1}{2}\sigma^{2}\right) \mp \frac{1}{2}\frac{M_{s}}{g_{s}}\,\sigma^{2}~.
\end{equation}
Here it is assumed that the recoil-velocities fluctuation strengths are the same for particle and antiparticle sectors (the asymmetric scenario has been studied in  \cite{Bernabeu:2006av}). As we can see, 
the local violation of Lorentz symmetry (LV) induced 
by the recoil velocities of the $D$-particles, induced a CPT violation too, since the dispersion relations 
between particles and antiparticles are different, generating a matter-antimatter lepton asymmetry
\begin{equation}\label{Leptonasym}
\ll n-{\overline{n}}\gg=g_{d.o.f.}\int\frac{d^{3}p}{(2\pi)^{3}}\ll[f(E)-f(\overline{E})]\gg
\end{equation}
where $f(E,\mu)=\frac{1}{{\rm exp}[(E-\mu)/T]\,\pm\, 1 }$, $E^{2}={\bf p}^{2}+m^{2}$, and $g_{d.o.f.}$ denotes the number of degrees of freedom of relativistic neutrinos. Assuming that $\sigma^{2}$ is constant
(independent of space and of the
(anti)neutrino energy), one can get $\Delta n_{\nu} \simeq \frac{g_{d.o.f.}}{\pi^{2}}\, T^{3}\left(\frac{M_{s}\sigma^{2}}{g_{s}\, T}\right)\,>\,0$. 
Notice that the CPT term, i.e., $-\frac{1}{2}\frac{M_{s}}{g_{s}}\sigma^{2}$,
in the dispersion relation of the neutrino comes with the right sign (\emph{``loss"}) guaranteeing the excess of particles over antiparticles. The resulting lepton asymmetry reads
\begin{equation}
\eta=\frac{\Delta n_{\nu}}{n_\gamma}\sim \frac{315}{2\pi^4}\frac{\text{GeV}}{T}\left(\frac{M_{s}}{\text{GeV}}\,
\frac{\sigma^{2}}{g_{s}}\, \right)~. \label{Ddl0}
\end{equation}
Observations of the CMB radiation \cite{[1]}, predictions of BBN \cite{[2]} (and the absence of intense radiation from matter–antimatter annihilation \cite{[3]}), implies that the observed baryon number asymmetry today is
\begin{equation}\label{etaBBN}
\eta = (6.04 \pm 0.08)\times 10^{-10}\,.
\end{equation}
Such a value remains constant from early times till today. 
For later reasons, one introduces the baryon density parameter $\eta_{10}$ defined as \cite{epjp52,epjp53}
\begin{equation}\label{eta10}
\eta_{10} \equiv 10^{10}\eta\equiv 10^{10} \frac{\Delta n_{\nu}}{n_{\gamma}}
\end{equation}
with $\eta_{10}$ to be determined. 
From (\ref{Ddl0}) and (\ref{eta10}) one gets
\begin{equation}\label{Msparam}
\frac{M_s}{\text{GeV}} \frac{\sigma^2}{g_s}=
10^{-13}\frac{2\pi^4}{315}\frac{T_{BBN}}{\text{MeV}}\, \eta_{10}
\end{equation}
where $T_{BBN}\sim 1$MeV is the temperature at which the BBN processes are effective.
%
%
%
\section{Primordial light element $\{ {}^4 He, D, Li\}$}
%
%
%
\par
\begin{figure}[btp]
 \centering
  \includegraphics[width=10.0cm]{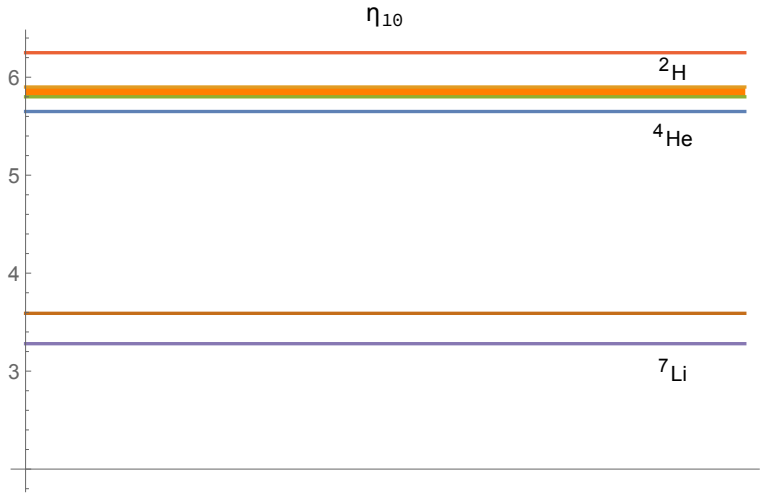}
  \caption{In this figure we report the bounds for $\{ {}^4 He$, $D, Li\}$, Eqs. (\ref{ZHe4}), (\ref{ZD}), and (\ref{ZLi}), respectively. The orange region is obtained from the overlapping of  (\ref{ZHe4}) and (\ref{ZD}). As we can see, the Lithium region, given by (\ref{ZLi}), does not overlap, so the Lithium problem cannot be solved or ameliorated by taking into account the spacetime foam model.}\label{eta10fig}
\end{figure}

\noindent
We will now derive the bound on the scale $M_s$  by analyzing the effects of the primordial abundances of light elements, i.e., Deuterium  ${}^{2}H$, Helium ${}^{4}He$, and Lithium ${}^{7}Li$, using the asymmetry given by (\ref{Ddl0}).
In this analysis, the baryon-antibaryon asymmetry, here indicated with $\eta_{10}$, plays a crucial role \cite{epjp52,epjp53}. Since we are interested in deviations from the standard cosmological model, hereafter we shall assume three generations of neutrinos so that we set $N_{\nu} =3$, which means $Z = 1$ (corresponding to the standard cosmological model) in the equations below.
We follow the Refs. \cite{batt,kavuk}. The relevant processes are here recalled: 
\begin{itemize}
\item $^{4}He$ {\it abundance} - The production of Helium $^{4}He$ is generated by the production of $^{2}H$ through a neutron and a proton. Consequently, the formed  Deuterium converts into $^{3}He$ and Tritium. 
The best fit of the primordial ${}^{4}He$ abundance is \cite{epjp57,epjp58}
\begin{equation}\label{Helium}
Y_{p} = 0.2485 \pm 0.0006 +   0.0016 \left[\left( \eta_{ 10} - 6\right) +100\left( Z-1\right)  \right]
\end{equation}
%
%
%
The standard result of BBN for the $^{4}He$ fraction is recovered for $Z=1$ and $\eta_{10} = 6$, so in General Relativity (GR) one gets $(Y_{p})|_{GR} = 0.2485 \pm 0.0006$.
However, observations of the Helium $^{4}He$ give
the abundance $0.2449 \pm 0.0040$ \cite{jcap2020}. So employing the observational constraint and the Helium abundance given in  (\ref{Helium}) for $Z=1$, we obtain
\begin{equation}
0.2449 \pm 0.0040 = 0.2485 \pm 0.0006 
+  0.0016 \left( \eta_{ 10} - 6\right)~. 
\end{equation}
From the above equations, one infers the constraint
\begin{equation}\label{ZHe4}
  5.65 \lesssim \eta_{10} \lesssim 5.9~.
\end{equation}
\item $^{2}H$ {\it abundance} -  Deuterium $^{2}H$ is produced by the reaction $n+p \rightarrow {}^{2}H +\gamma$.
The best fit gives the  Deuterium abundance \cite{epjp52}
\begin{equation}\label{16}
y_{D p} = 2.6(1 \pm 0.06) \left( \frac{6}{\eta_{ 10} - 6 (Z-1)}\right)^{1.6}~.
\end{equation}
The values $Z=1$ and $\eta_{10} = 6$ yield the result in GR, thus the Deuterium abundance will be $y_{D p} |_{GR} = 2.6 \pm 0.16$.
Equation (\ref{16}) and  the observational constraint on deuterium abundance $y_{D p} = 2.55 \pm 0.03$ \cite{jcap2020} give (for $Z=1$)
\begin{equation}
2.88 \pm 0.22 = 2.6 (1 \pm 0.06) \left( \frac{6}{\eta_{ 10}}\right)^{1.6}~.
\end{equation}
One then gets the constraint
\begin{equation}\label{ZD}
 5.88626 \lesssim \eta_{10} \lesssim 6.25264~.
\end{equation}
It is noteworthy that such a constraint partially overlaps the helium abundance (\ref{ZHe4}).
\item $^{7}Li$ {\it abundance} - 
The parameter $\eta_{10}$, defined in (\ref{eta10}), although  successfully fits the abundances of $D$ and ${}^4He$, it does not fit the observations of $^{7}Li$. 
This is referred in literature as {\it the Lithium problem} \cite{theory}). 
The ratio of the expected value of ${}^{7}Li$ abundance in GR and the observed one is in the range \cite{theory,theory42}    
\begin{equation}\label{LitabGR}
\frac{Li|_{GR}}{Li|_{obs}}  \in [2.4-4.3]~.
\end{equation}
The numerical best fit for $^{7}Li$ abundance is (for $Z=1$) \cite{epjp52}
\begin{equation}
y_{Li} = 4.82 (1 \pm 0.1)\left[\frac{\eta_{ 10} - 3 (Z-1)}{6} \right]^{2} = 4.82 (1 \pm 0.1)\left[\frac{\eta_{ 10}}{6} \right]^{2}~.
\end{equation}
Employing the observational constraint on Lithium abundance, i.e., $y_{Li} = 1.6 \pm 0.3$ \cite{jcap2020}, one gets the constraint
\begin{equation}\label{ZLi}
 3.28457 \lesssim \eta_{10} \lesssim 3.59177~.
\end{equation}
It is evident that such range of values does not overlap with the constraints on ${}^2 H$ abundance, i.e., Eq. (\ref{ZD}), and on ${}^4He$ abundance, i.e., Eq. (\ref{ZHe4}).
\end{itemize}
The constraints derived for the three abundances are reported in Fig. \ref{eta10fig}. It is obvious that the overlapping ranges of ${}^2 H$ and ${}^4 He$ correspond to the value $\eta_{10}\sim 5.9$ (orange region).
This value does not overlap with the ${}^7 Li$ range, which means that the Lithium problem cannot be solved in the framework of the spacetime foam. However, this is not a conclusive result since the eventual modifications of Einstein's equations have not been considered in the present paper, as well as the possibility that the quantum fluctuation parameter $\sigma^2$ is not universal.

%
%
Inserting the overlapping value of  $\eta_{10}$ into (\ref{Msparam}) for $T_{BBN}\lesssim 1$MeV and using 
spacetime foam parameter $M_s/g_s\sim M_P$ ($M_P\sim 10^{19}$GeV is the Planck mass) \cite{Nick2}, one obtains
\begin{equation}\label{BBN_bound}
\frac{M_s}{g_s}\sigma^2 \sim  3.6 \times 10^{-13} \text{GeV} \quad \to \quad \sigma^2 \lesssim 10^{-22}~.
\end{equation}
Therefore, we have inferred the upper bound on the dimensionless stochastic variable $\sigma^2$, which expresses the fluctuations of the recoil velocity of the $D$-branes.
%
%
%
%
%
%
%
%
%
%
%
%
%
%
%
\section{Conclusions}
\label{sec:3}
%
%
%
\par\noindent
The BBN era, which occurred during the hot and expanding early Universe, has left an observable imprint in the abundance of primordial light elements. 
Precision observations and high-accuracy predictions of these elements provide an important test of the standard cosmological model (based on General Relativity)  and allow probing of non-standard cosmological and particle physics scenarios. 
In this framework, we have used the BBN sensitivity to obtain a bound on the dimensionless stochastic variable $\sigma^2$ expressing the fluctuations of D-branes recoil velocity. Results give $\sigma^2 \lesssim 10^{-22}$ for $M_s/g_s \sim M_P$. A follow-up of the present work is to investigate physical scenarios related to spacetime foam where the BBN constraints are properly taken into account, or consider the general case in which the quantum fluctuation parameter $\sigma^2$ is not universal \cite{nickEPJC}.  

%
%
%
\acknowledgements
%
%
%
\par\noindent
The authors would like to thank N. Mavromatos for useful correspondences and fruitful comments.
This work was supported by the Natural Sciences and Engineering Research Council of Canada. 
%
%
%

%
%
%
%
\end{document}